\documentclass[
 reprint,
 amsmath,amssymb,
 aps,
floatfix,
]{revtex4-2}

\usepackage{graphicx}
\usepackage{bm}
\usepackage{gensymb}
\usepackage{upgreek}
\usepackage{mathptmx}
\usepackage{array}
\usepackage[english]{babel}
\usepackage{xcolor}

\begin{document}

\preprint{APS/123-QED}

\title{Coffee-Ring Deposits in Concentrated Suspensions of Anisotropic Colloids}

\newcommand{\ryker}[1]{\textcolor{violet}{\textbf{Ryker:} #1}}

\newcommand{\mmd}[1]{\textcolor{blue}{\textbf{Michelle:} #1}}


\author{Samuel S.\ Nielsen}
\altaffiliation[Currently at ]{Department of Physics, Brandeis University}
\affiliation{Department of Physics and Astronomy, Northwestern University}

\author{Michelle M.\ Driscoll}
\email{michelle.driscoll@northwestern.edu}
\affiliation{Department of Physics and Astronomy, Northwestern University}

\author{Brian C.\ Seper}
\affiliation{Department of Physics and Astronomy, Northwestern University}

\author{Ryker Fish}
\affiliation{Applied Mathematics and Statistics, Colorado School of Mines}

\author{Brennan Sprinkle}
\affiliation{Applied Mathematics and Statistics, Colorado School of Mines}

\begin{abstract}
Evaporating a droplet containing dispersed colloids leaves behind a dried deposit whose shape is determined by capillary flows and the resulting particle transport. The classical coffee-ring effect occurs when an outward radial flow drives particles toward the droplet's contact line as the droplet evaporates, resulting in uneven deposition. This deposition is often studied in dilute concentration regimes where, hydrodynamically, the effects of particle shape are unimportant. As particle concentration increases, it is expected that particle anisotropy should play a larger role in modifying transport and potentially suppressing coffee-ring formation. We present experiments isolating the effects of particle shape, concentration, and density, as well as solvent temperature, on the geometry of the ring deposit. By analyzing the deposits using surface profilometry to more accurately characterize ring widths, these experiments show that coffee-ring formation is independent of particle anisotropy and is instead controlled by the ratio of the particle sedimentation velocity to the velocity of the droplet's evaporating air-water interface. Hydrodynamic simulations support this finding by providing quantitative estimates for bulk sedimentation velocity. Together, these results offer a unified picture of how multiple physical parameters determine coffee-ring geometry with direct implications for suppressing uneven deposition in practical applications. 
\end{abstract}

\maketitle

\section{Introduction}

During the evaporation of a droplet of a colloidal suspension, a pinned contact-line induces a radial outward flow to replenish the water lost to evaporation along the droplet's perimeter. This results in the suspended colloids being transported to the edge of the droplet where they pack together and leave a high concentration of colloids along the circumference and a low concentration of colloids near the center. This effect, first studied in detail by Deegan~\cite{deegan1997capillary}, is known as the coffee-ring effect. Many applications of colloidal suspensions, such as coatings~\cite{giorgiutti2018drying}, spray-dried liquid food~\cite{food,giorgiutti2018drying}, and inkjet printing~\cite{inkjet} would be made more efficient if smoother deposits were left upon evaporation. As such, much of the research concerning the coffee-ring effect is focused on its suppression. The coffee-ring effect can be suppressed by interface capture. Interface capture can occur in systems of colloidal nanoparticles via absorption to the interface which is driven by surface affinity; often these particles nucleate into ordered, two-dimensional structures along the interface~\cite{bigioni,ALMILAJI2018234,silicaNano}. Interface capture can also be attributed to a kinetic effect where the evaporating interface descends more quickly than solute particles can organize into the coffee ring pattern, thus halting its formation~\cite{interfaceassemblyreview,yunker2011suppression,PolymerCoat}. Previous work has shown that parameters which impact the final deposit include temperature~\cite{li2015coffee}, surface treatments of both the particle~\cite{PolymerCoat} and substrate~\cite{oilysurface,Boulogne2017CREonDryandWetSurfaces}, particle density~\cite{DensityDependence}, and particle anisotropy~\cite{yunker2011suppression,dugyala2014control,2025rodscolor}. Previous explorations of particle anisotropy have revealed its potential for tuning deposit formation~\cite{yunker2011suppression,dugyala2014control}, but differences in the suspending medium and particle surface chemistry make it challenging to draw conclusions the effect of anisotropy on the structure of the deposit.

\begin{figure*}[htb]
\centering
  \includegraphics[height=10cm]{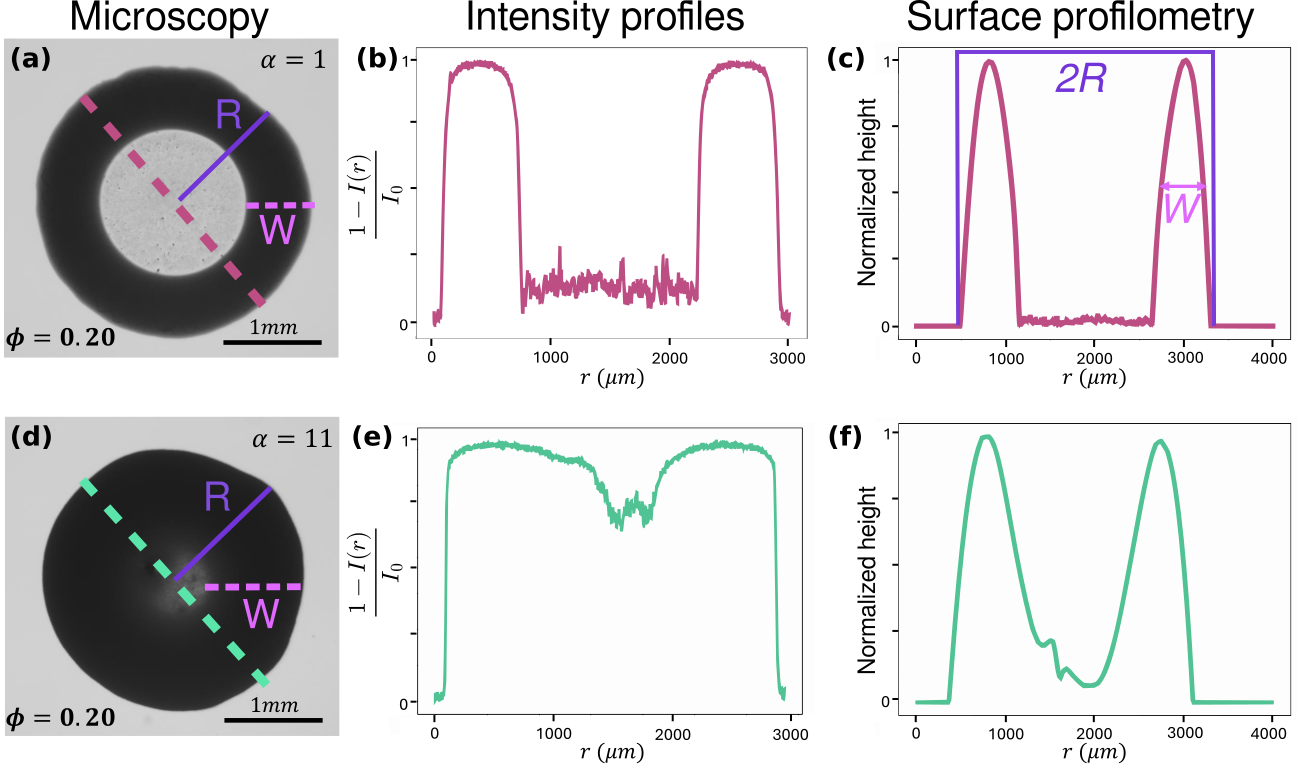}
  \caption{\textbf{Surface profilometry captures ring deposit geometry.} Bright-field microscopy images of deposits formed by evaporating droplets with silica particles of aspect ratio $\alpha$ for (a) $\alpha=1$ and (d) $\alpha=11$, both at $\phi=0.20$. The coffee ring is characterized by a ring width, $W$, and a pattern radius, $R$; these are labeled in (a), (c), and (d). (b) and (e) show the normalized intensity profiles $\left[1-I(r)\right]/I_0$ taken along the dashed lines, where $I_0$ is the intensity of the white background. (c) and (f) show the result of surface profilometry along the dashed lines.}
  \label{fig:figure1}
\end{figure*}

In this work we thoroughly examine the effect particle anisotropy has on ring formation by evaporating suspensions of silica colloids in water. All anisotropic particles are synthesized using the same method, ensuring that the surface chemistry is unchanged for all aspect ratios. Previous experimental~\cite{cationspolydispersed} and theoretical~\cite{popov2005evaporative} investigations into the coffee-ring effect have focused on volume fractions of colloids $\phi < 0.1$~\cite{BigBoiParticlesMakeEvenDeposits2021,Boulogne2017CREonDryandWetSurfaces}. We measure the ring width over a broad range of volume fractions up to $\phi = 0.35$ and demonstrate that the effects of anisotropy are not concentration dependent. Surface profilometry is used to make accurate measurements of the deposit's shape and characteristics, even at high concentrations where microscopy techniques can fail to capture ring structure. We also obtain profiles for both silica and polystyrene suspensions evaporated at room temperature (20 \degree C) as well as in a 90 \degree C oven, and show that the difference in velocity between the descending air-water interface and the sedimenting colloids is likely responsible for the suppression of coffee ring formation. 

\section{Materials and Methods}
\subsection{Synthesis of Silica Particles and Sample Preparation}

\begin{figure}[h]
 \centering
 \includegraphics[height=8cm]{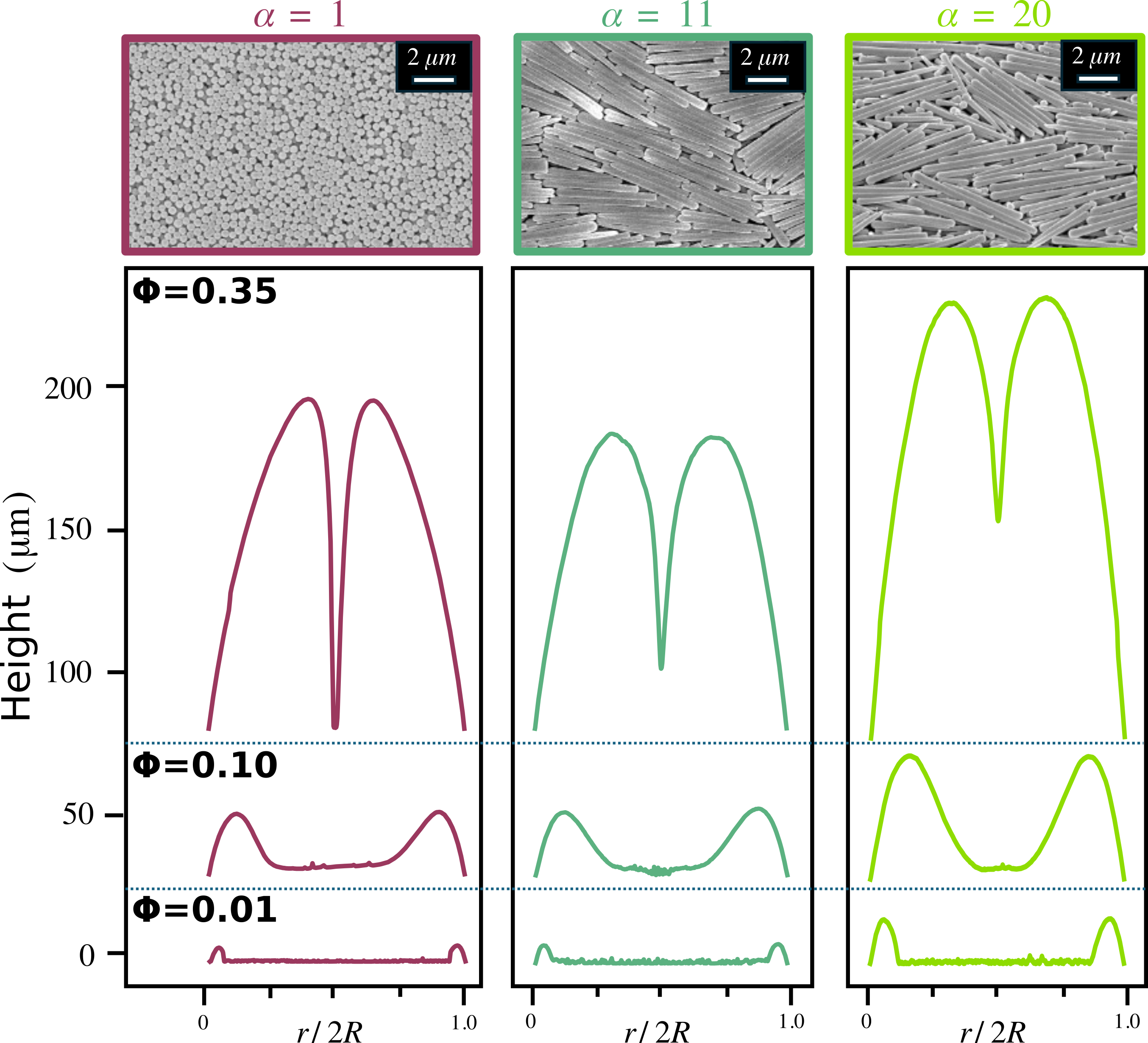}
 \caption{\textbf{The volume fraction of the droplet dictates the geometry of ring deposit independent of particle aspect ratio}. Profilometry scans for $\alpha=1,11$ and $20$ are shown at volume fractions $\phi = 0.01$, $0.1$ and $0.35$. Profiles are vertically offset for clarity, and normalized horizontally by their size $2R$. Increasing volume fraction causes the ring width to increase, while the aspect ratio has no appreciable effect on the nature of the profile. Above the profile scans are representative SEM images for each aspect ratio. (SEM scans are of independent samples, not of the dried deposits.)}
 \label{fig:2}
\end{figure}

Charge-stabilized silica spheres (diameter = 300 $\pm$ 30 nm, 830 $\pm$ 20 nm) were fabricated using the St\"{o}ber synthesis technique, \cite{stober1968controlled, zhang2009hollow} and rod-shaped silica particles were created using a one-sided emulsion-nucleation reaction~\cite{kuijk2011synthesis}. By varying the temperature of the reaction, as well as the age of the reagents, different aspect ratio rods were produced. For the rods used in this study, reaction temperatures of 18, 27, 31, and 31 \degree C yielded particles of aspect ratio $\alpha = 3\pm 0.5,$ $7 \pm 1.4$, $11\pm1.4$, and $20\pm2.3$, respectively. Synthesis at 31 \degree C yielded both aspect ratio 11 and 20 rods due to a higher ammonia concentration, which is known to produce shorter rods ~\cite{kuijk2011synthesis}. All particles were centrifuged and resuspended in ethanol 4 times in order to clean the particles and their medium. Following this cleaning particles were resuspended in MilliQ water for use in experiments. To reduce polydispersity, particle suspensions were centrifuged at $700\textit{g}$ for 15 minutes and their supernatants were removed. Leaving the suspensions at $1\textit{g}$ for 10 minutes allowed for larger than desired particles to be extracted from the bottom of the solution. Cleaned particles were imaged using a Hitachi S4800 Scanning Electron Microscope (see Figure \ref{fig:2}). The diameters and aspect ratios of the particles were characterized from these images using the open source software ImageJ \cite{imageJ}. Error bars reported on particle size and aspect ratio come from the standard deviation from a minimum sample of 20 particles. 

The concentration of suspensions were determined by measuring the mass fraction of particles and converting those to volume fraction. The wet and dry mass of 80 $\mu$L of solution was recorded on a balance, and the ratio of the dry mass to the wet mass is the mass fraction of the particles. With this, the volume fraction of the suspension was calculated directly as the densities of silica rods and spheres are known as $1.9 g/cm^3$\cite{kuijk2011synthesis} and $2.2 g/cm^3$, respectively \cite{spheresynth1}\cite{spheresynth2}. We note the difference in density between particles of the same material are due to differences in the synthesization processes, however surface chemistry remains unchanged. Solutions were then concentrated or diluted to the desired volume fraction in order to perform experiments. To ensure droplets had the desired concentration, the concentration of the solution the droplet was taken from was calculated using the above procedure at the start and end of the experiments, and the difference was never larger than $1\%$. 

To test the role of particle density, carboxylated polystyrene spheres (diameter = 288 nm) were procured from Bangs Laboratories Inc. (Fishers, IN) to compare against experiments performed used higher-density silica spheres. The polystyrene spheres were cleaned 8 times via centrifugation and re-suspension in MilliQ water to remove surfactants from the suspending medium.

\subsection{Substrate Preparation and Experimental Procedure}

 Substrates were prepared as follows. Microscope slides were soaked with a 2.5 M solution of NaOH in a 60:40 water to ethanol mixture (by volume) for 2 hours to remove organic contaminants. The slides were then thoroughly rinsed with deionized water for 10 minutes, dried using compressed air, and used immediately for experiments. In order to form droplets for these experiments, 1 $\upmu$L of a colloidal suspension of a known volume fraction was pipetted onto the cleaned glass slide. The outlined procedure resulted in droplet radii between 1.2 mm and 2 mm, depending on the concentration of colloids in the droplet. Assuming the drop shape to be a spherical cap, the droplet radii is used to calculate the contact angle as being between 7\degree and 35\degree. We attribute the high variation in contact angle to inconsistency in substrate surfaces after cleaning. Additionally, differences in evaporation time was not seen to significantly affect our measurements as drops dried at 75\% relative humidity produced profiles consistent with those seen at much lower humidities. The drying process was imaged from below using a microscope setup built in the lab. Most experiments were performed in ambient conditions (T $\approx$ 20 $\pm$ 2 \degree C, 10$\%$ $\leq$ relative humidity (RH) $\leq$ 50 $\%$). The variation in RH represents seasonal variation in the laboratory. Some experiments were performed with high relative humidity (RH $\geq$ 75$\%$) or performed on preheated substrates in an oven set to 90 \degree C.

In order to directly compare to results from the literature \cite{li2015coffee}, commercially acquired polystyrene particles and synthesized silica spheres (diameter = 300 $\pm$ 30 nm) were dried on a 5mm x 7mm silicon wafer acquired from Ted Pella (Redding, CA). The wafers were rinsed briefly in deionized water, then acetone, and then isopropyl alcohol and dried using compressed air. This procedure resulted in average droplet radii of 1.4 $\pm$ 0.25 mm, from which we calculated a contact angle of 25.6 $\pm$ 12.8\degree.

\subsection{Profilometry Measurements and Analysis}
The geometry of the dried deposits were characterized using a Dektak 150 Surface Profilometer. Surface profilometry was chosen to ensure that rings would be accurately observed, even at high volume fractions where microscopy might mischaracterize the deposit due to image saturation. The value of profilometry characterization is clearly illustrated in Figure \ref{fig:figure1}. 
For particles with $\alpha=1$, the coffee ring is visible in both the intensity profile and surface profilometry. However, for $\alpha=11$, the intensity profile appears mostly even due to saturated absorbance while profilometry reveals a distinct ring. To ensure the deposits were not being damaged by the instrument, scans were repeated several times over the same path and the profiles were found to be identical. Representative profile scans are shown in Figures \ref{fig:figure1} and \ref{fig:2}.

Profilometry data was analyzed by first finding the location and height of the peaks of the coffee ring. The full-width-half-max (FWHM) were obtained by finding the first point below the half-maximum on each side of the peaks. The radius of each droplet was found by locating local maxima in the profile's second derivative, corresponding to the edges of the droplet. The width and radius are shown on a sample profile in Figure \ref{fig:figure1}. For experiments with $\phi = 0.35$, there are a small number of samples whose FWHM is not well defined; this can be seen in Figure \ref{fig:2} where the lowest point in the center of the profile does not reach the half-max when $\alpha=20$. In this case, the width is calculated at the center minimum. We note that such samples are uncommon, and the difference in widths calculated using this methods have only small differences from samples where the FWHM exists.

\subsection{Simulations}

Simulations were performed for spherical and rod-shaped particles to obtain numerical estimates for bulk suspension sedimentation velocity in the overdamped (Stokes) regime. A suspension of particles is modeled by simulating in a triply periodic box of side length $L_b = 7500$ nm. Hydrodynamic interactions between particles are included using the PSE method from the software library libMobility, which is a fast implementation of the periodic Ewald sum for Rotne-Prager-Yamakawa kernels \cite{pse, libmobility}. To improve hydrodynamic resolution and allow for non-spherical particles, one physical particle is discretized using many smaller ``blobs'' that are used to calculate hydrodynamic interactions~\cite{delong2014}. Spherical particles are discretized using icospheres (geodesic polyhedrons) and rod-like particles of different aspect ratios are discretized as spherocylinders; see insets on Figure~\ref{fig:vel_vs_phi}c) for illustrations of these discretizations. We utilize the rigid multiblob method which allows hydrodynamics to be calculated blob-wise but provides the resulting particle-level velocity while constraining each body to remain rigid~\cite{rigidMultiblobOriginal, rigidMultiblobReview}.

During each simulation, we applied a gravitational force $F_g = mg$ to the center of mass of each particle, where $m$ is the excess mass of the particle and $g$ denotes acceleration from gravity. Due to the type of regularized immersed boundary method used to compute hydrodynamics, it is possible for particle overlaps to occur. To prevent these overlaps, a soft repulsive (steric) potential is imposed between blobs. We use the potential from Eqn.\ 17 from Fish \textit{et al.}~\cite{libmobility} with $b=0.1r$ and $U_0=4 mgL_p$, where $r$ is the radius of a blob and $L_p$ is some length scale associated with the particle geometry. We found that $L_p = V_p^{1/3}$, where $V_p$ is the volume of the particle, successfully eliminated overlaps sufficiently without being unnecessarily large. Dynamics are performed using an Euler method, and the computational methodology was validated by confirming the velocity of a single sedimenting particle matches the analytic expression for Stokes drag (including the appropriate periodic correction)~\cite{pelaezSpectralSolverOscillatory2025}.

Initial configurations for each volume fraction and particle shape/size were created for spherical and rod-shaped particles using codes from Skoge \textit{et al}. and PACKMOL, respectively~\cite{skogeSpherePacking2006, packmol, packmol_packing}. Due to the steric repulsion forces between particles and imperfect packings, simulations needed to be equilibrated to prevent spurious velocity fluctuations caused by overlapping particles at initialization. This was done by first simulating using a small timestep $\Delta t = 0.25 \tau_U$, where $\tau_U = \frac{6\pi\eta r^2 b}{U_0}$ is the blob-level steric repulsion timescale, until the largest difference in average suspension velocity over a window of $10 \tau_U$ was less than 1\% of the mean over the same window. After a packing was equilibrated to steric interactions, the simulation could be run with a larger timestep of $\Delta t \propto \tau_g$, where $\tau_g = \frac{r}{M_0 mg}$ is the timescale for a particle to sediment one blob radius when $M_0$ is the mobility coefficient for the entire particle. The exact stable timestep varies per simulation, but in general a smaller multiple of $\tau_g$ was needed for denser suspensions. We use $M_0 = \frac{1}{6 \pi \eta a}$ for spherical particles and $M_0 = \frac{\ln\left[L/(2a)\right]}{4 \pi \eta L}$ for rod-shaped particles, where $a$ is the particle radius and $L$ is the rod length~\cite{batchelorSlenderbodyTheoryParticles1970}. Each simulation was ran until the mean suspension sedimentation velocity was converged, and simulation convergence was assessed using forward/reversed convergence plots from Klimovich et al.~\cite{klimovichGuidelinesAnalysisFree2015}. A simulation was run until the cumulative mean of the last 50\% of the data was within half of a standard deviation of the mean of the entire dataset after any apparent equilibration period was removed by visual inspection. 

\section{Results and Discussion}
\begin{figure}
    \centering
    \includegraphics[height=7cm]{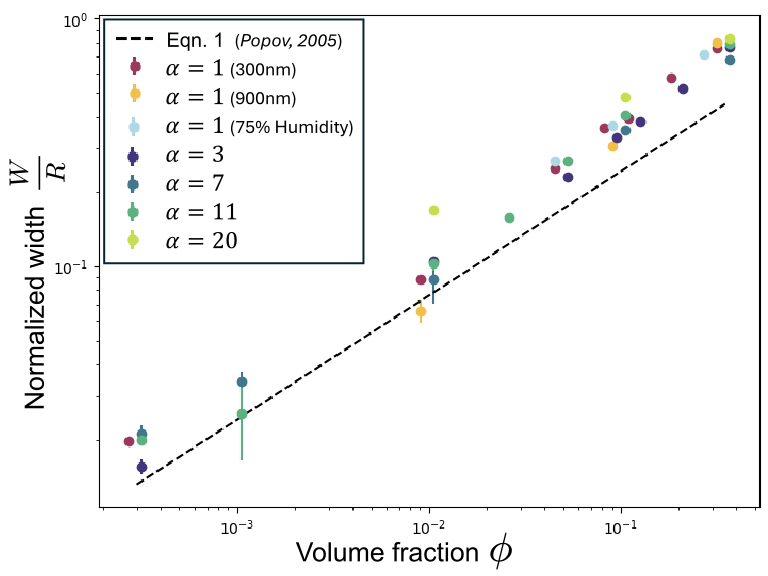}
    \caption{\textbf{Coffee-ring geometry does not depend on particle anisotropy.} 
    Normalized widths of the ring shaped deposits formed by droplets containing silica colloids of aspect ratio $\alpha$ are plotted against initial volume fraction. The dashed line is equation~\ref{eq:popov} using the random close packing value for spheres, $\phi_{\text RCP}(1)=0.64$.}
    \label{fig:figure3}
\end{figure}

Figure~\ref{fig:figure3} shows the ring widths, $W$, normalized by droplet radius, $R$, plotted against the volume fraction for all tested aspect ratios and conditions. We see that the data collapses nicely to a power-law scaling. A prediction for this dependence of the width on particle concentration was made by Popov~\cite{popov2005evaporative}: 

\begin{equation}
    \frac{W}{R} = 0.609\sqrt{\frac{\phi}{\phi_{\text RCP}(\alpha)}},
    \label{eq:popov}
\end{equation}

where $\phi$ is the volume fraction of colloids in the droplet and $\phi_{\text RCP}(\alpha)$ is the maximum random close packing volume fraction for aspect ratio $\alpha$, e.g. for spherical particles, $\phi_{\text RCP}(1) \approx 0.64$. Equation~\ref{eq:popov} was derived assuming a low concentration ($\phi < 0.1$) and did not take into account the particle composition, the cross-sectional geometry of the ring, or hydrodynamic interactions between particles. Surprisingly, Figure \ref{fig:figure3} shows that our data matches well with the predicted scaling relationship, even for  $\phi > 0.1$. Given the assumptions under which equation~\ref{eq:popov} was derived, we do not expect the prefactor to be an exact prediction for high particle concentrations. Additionally, the width predicted by Popov is not necessarily the FWHM as we have measured. These discrepancies are seen in Figure \ref{fig:figure3} as the dashed line slightly underestimates the ring width.

Equation \ref{eq:popov} also tells us that the width should be inversely proportional to the square root of $\phi_{\text RCP}(\alpha)$. The maximum packing fraction of spherocylinders is known to decrease rapidly with aspect ratio, although it stays near that of spheres until around $\alpha=10$ \cite{PackingFrac}. This is consistent with our measurements where data points for $\alpha > 10$ tend to sit above the rest of the dataset. We also note that our results show strong agreement with the square-root scaling in the high concentration limit which was not considered in Popov's work \cite{popov2005evaporative}.

\begin{figure*}[ht]
    \centering
    \includegraphics[height=4.1cm]{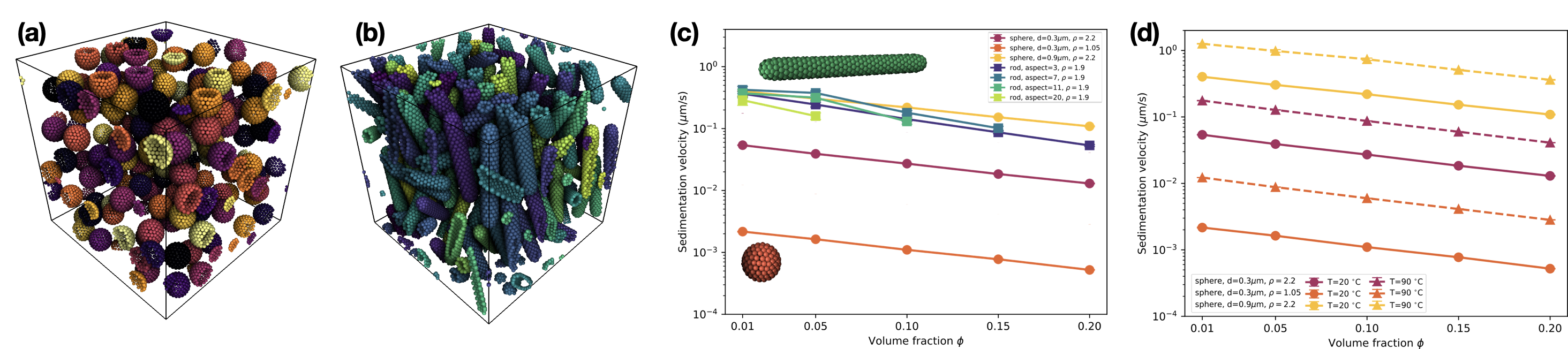}
    \vspace{-3mm}
    \caption{\textbf{Quantification of sedimentation velocity via simulation as a function of packing fraction, particle shape, and temperature.} Example simulation runs for (a) spheres and (b) rods are shown for $\phi$ = 0.15. (c) Mean suspension sedimentation velocity vs.\ $\phi$ for our range of particle shapes and sizes. Insets illustrate the discretization used for a 300 nm diameter sphere and an a rod of aspect ratio 11 (diameter 384 nm, length 4224 nm). (d) Mean suspension sedimentation velocity vs.\ $\phi$ for spherical particles at 20 \degree C (circles, solid lines) and 90 \degree C (triangles, dashed lines). For (c, d), error bars given by standard deviation of sedimentation timeseries are smaller than marker size.}
    \label{fig:vel_vs_phi}
\end{figure*}

Given that the coffee-ring effect persists over all tested concentrations and aspect ratios, we observe that particle shape alone does not change the characteristics of the deposit or suppress the coffee-ring in systems containing silica colloids. Previous works using polystyrene colloids has proposed that the coffee-ring effect is suppressed when colloids are captured by the air-water interface during evaporation \cite{yunker2011suppression, li2015coffee} or when particles quickly sediment and stick onto the substrate~\cite{DensityDependence}. This study, as well as others which observed rings using ellipsoidal particles~\cite{dugyala2014control}, used silica or hematite colloids, both of which have a much higher density relative to water than polystyrene. The key difference here is the sedimentation rate, which is determined by the colloid's density and size. Balancing the gravitational force with Stokes' drag force, we can write the terminal velocity of a sedimenting spherical
\begin{equation}
    v_{p}^{sph} = \frac{2 \Delta\rho g a^2}{9 \eta},
    \label{eq:sediment_sphere}
\end{equation}
or rod-shaped
\begin{equation}
    v_{p}^{rod} = c_0\frac{ \ln \left[L/(2a)\right]\Delta\rho g a^2}{4 \eta },
    \label{eq:sediment_rod}
\end{equation}
colloid \cite{batchelor2000introduction, batchelorSlenderbodyTheoryParticles1970},
where $\Delta\rho$ is the difference in density between the colloid and the suspending medium, $g$ is the acceleration due to gravity, $a$ is the radius of the sphere/rod, $L$ is the length of the rod, $\eta$ is the dynamic viscosity of the liquid, and $c_0$ is a coefficient of order one that depends on the orientation of the rod. 

For polystyrene, the difference in density of the particles from that of water is 0.05 g/cm$^3$ while for silica spheres it is 1.2 g/cm$^3.$ At 22 \degree C,  equation~\ref{eq:sediment_sphere} gives that the sedimentation velocity of a silica colloid is 0.234 $\mu$m/s and a polystyrene colloid is 0.009 $\mu$m/s. There are in general, not purely analytic expressions for the sedimentation of suspensions, and thus we employ numerical simulations to compute the sedimentation velocity in the suspensions. We performed simulations of sedimenting suspensions of anisotropic particles including overdamped hydrodynamic interactions and contact forces, see details in Methods and Materials. The simulations are done in periodic boxes to model bulk suspensions up to $\phi = 0.2$. As seen in Figure \ref{fig:vel_vs_phi}, the sedimentation velocity varies by less than an order of magnitude as $\phi$ is increased, as predicted and measured by previous works~\cite{brzinski2018observation, richardson1997sedimentation, batchelor1972sedimentation}. However, there is a large difference in sedimentation velocities of polystyrene vs.\ silica that persists for all $\phi$.  Thus, we expect (and observe) that for a range of interface velocities, the descending air-water interface can capture the polystyrene particles but moves too slowly to capture the silica particles. Once the particles are captured on the surface, they assemble into a structure which grows towards the center of the droplet and accumulates other surface-captured particles before they can reach the edge~\cite{PolymerCoat}. This results in a more evenly dried deposit. We note that quantitative comparisons to the interface velocity are difficult to make given our imaging modality and fluctuations in evaporation times. The observed evaporation time ranged from 1 to 15 minutes depending on the ambient humidity and the particle concentration, although the range of evaporation times restricts to between 1 and 4 minutes if we exclude the cases where RH $=75\%.$ Despite this variation, our measured widths (and lack of even deposition) reveal that significant interface capture does not occur in samples containing silica particles as we only observed ring-shaped deposits. 

\begin{figure*}[t]
    \centering
    \includegraphics[height=8cm]{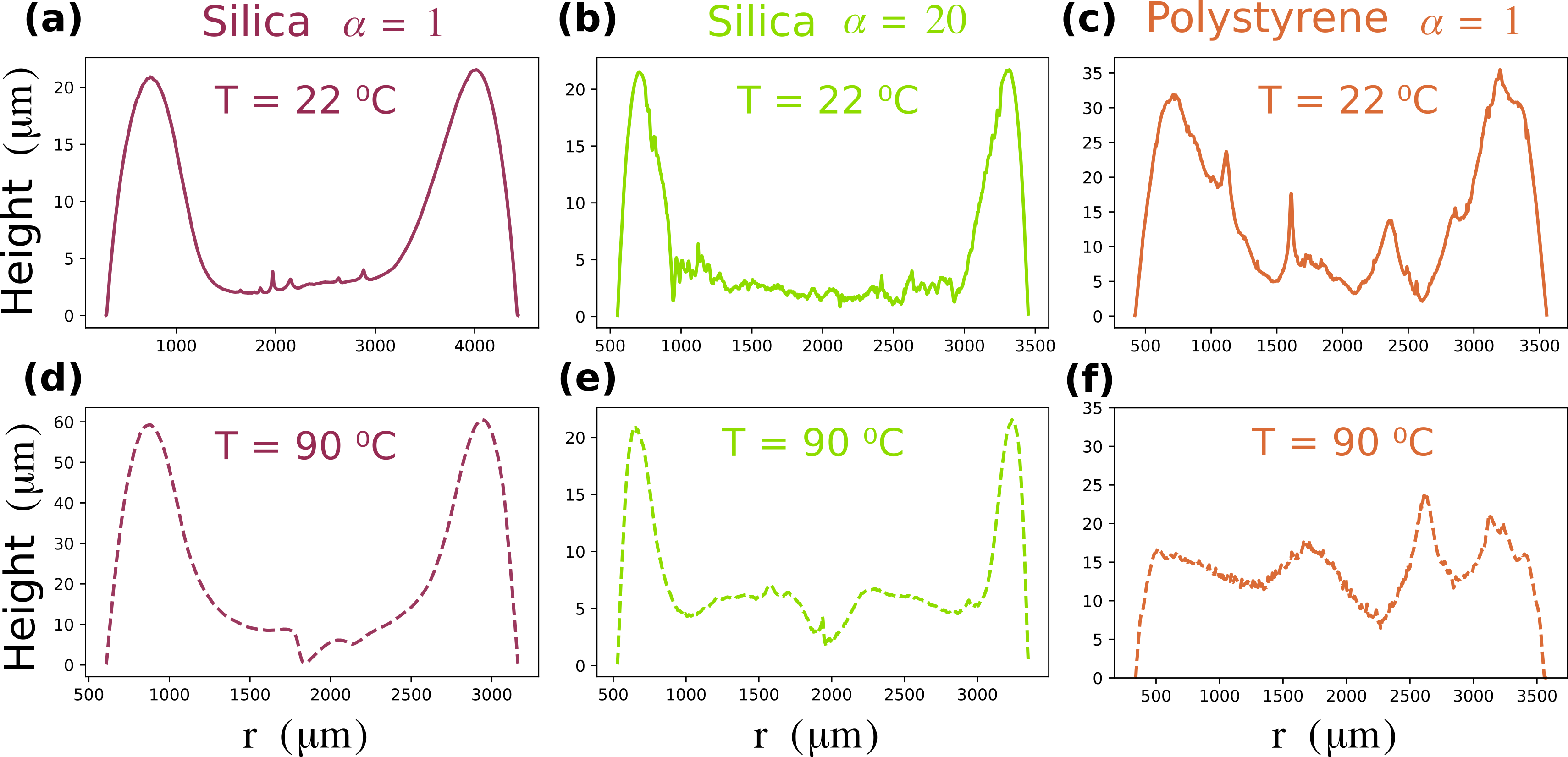}
    \vspace{-2mm}
    \caption{\textbf{Deposits become smoother when particles are more easily captured by the interface.} Deposits formed from $\phi =0.1$ silica spheres (a,d), $\phi = 0.1$ silica rods with  $\alpha = 20$ (b,e), and $\phi = 0.13$ polystyrene spheres (c,f) at 22 \degree C (a-c) and 90 \degree C (d-f). Profiles exhibit a clear ring shape except for polystyrene evaporated at 90 \degree C  in (f). At higher temperatures, the air-water interface moves fast enough to capture polystyrene during evaporation, resulting in a more uniform deposit. Denser silica particles sediment too quickly to be captured by the interface before being carried to the droplet edge.}
    \label{fig:figure4}
\end{figure*}

\begin{figure*}[th]
    \centering
    \includegraphics[height=8cm]{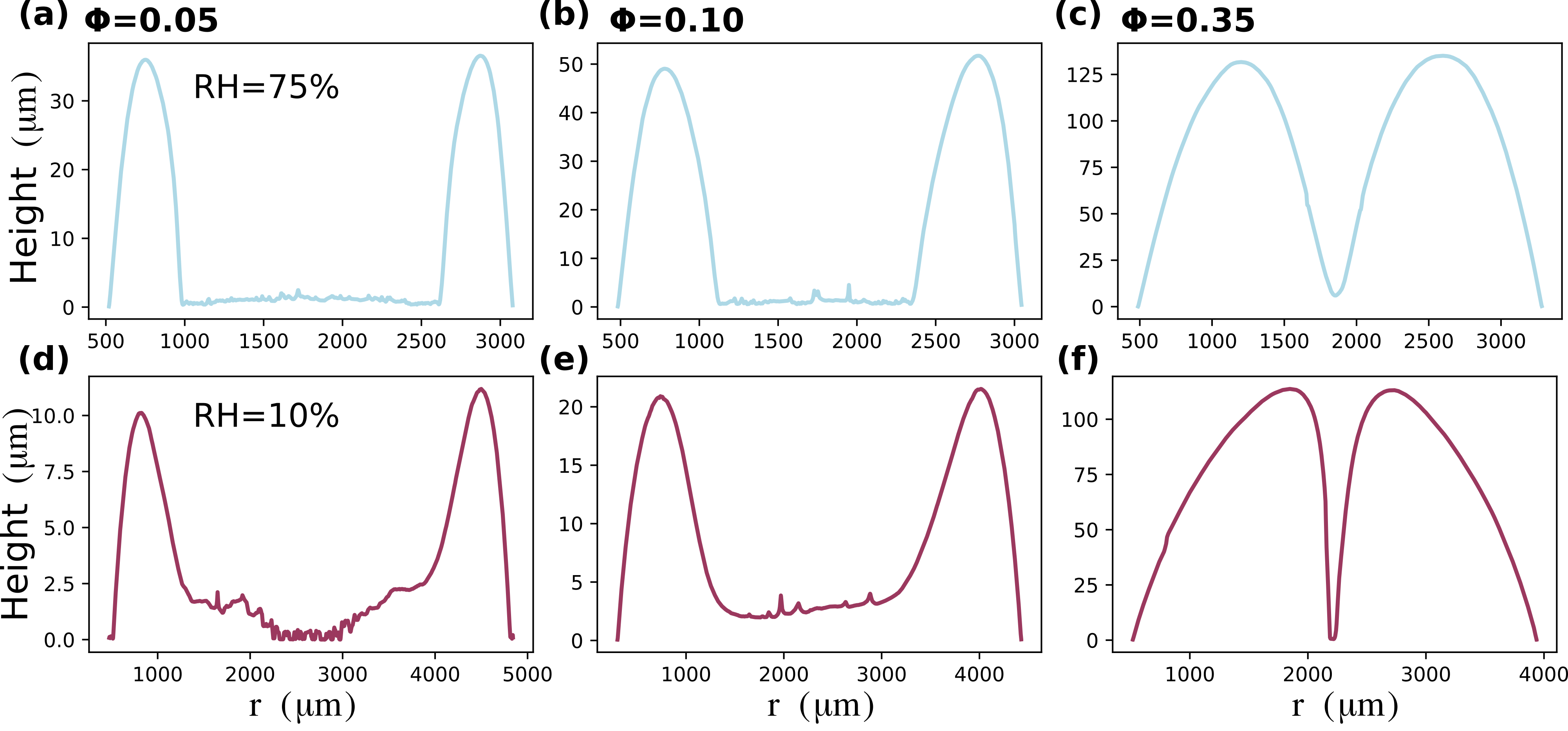}
    \vspace{-2mm}
    \caption{\textbf{Increasing the relative humidity does not effect ring formation.} Profilometry formed by 300nm silica spheres at 75\% relative humidity (a-c), and 10\% relative humidity (d-f), at $\phi = 0.05$ (a,d), $0.1$ (b,e), and $0.35$ (c,f). The measured width of the coffee ring does not change with increased relative humidity in this system of silica colloids.}
    \label{fig:RHprofiles}
\end{figure*}

The velocity of the descending interface, $v_i$ is controlled by the evaporation rate. To examine how changes in $v_i$ alter deposit patterns, we obtained profiles for silica and polystyrene sphere suspensions dried at 22 \degree C and at 90 \degree C. Figure \ref{fig:figure4} shows these profiles, from which we observe that the deposition of polystyrene particles can be smoothed by increasing the system's temperature, in agreement with measurements reported by Li et al.~\cite{li2015coffee}. Silica particles, on the other hand, continue to form rings even at high temperatures. This is consistent with our results that variation of RH from 10--75\% did not alter the geometry of the observed pattern (see Figures~\ref{fig:figure3} and \ref{fig:RHprofiles}). While temperature can drive strong Marangoni flows in some systems (for example for suspensions in organic solvents), it has been shown that in aqueous systems these effects are quite weak due to hard-to-suppress surface contamination~\cite{hu2006marangoni, li2015coffee}.  In our system, the effect of temperature on pattern geometry can be understood by examining the dependence of $v_p$ and $v_i$ on temperature, as illustrated in Figure \ref{fig:vel_vs_phi}d). As the temperature increases both velocities increase but by vastly different amounts: $v_p$ increases by a factor of 3-5 (governed by $\Delta \rho$) due the reduction in $\eta$, while the interface velocity increases by over an order of magnitude~\cite{li2015coffee}, as expected from the large change in vapor pressure from 20--90 \degree C. This means that the interface is faster relative to the colloids at higher temperatures than it is at lower temperatures, enhancing adsorption to the interface.
Figures \ref{fig:figure3} and \ref{fig:figure4} demonstrate that particle shape alone cannot control the deposition pattern, nor can temperature. Instead, it is a combination of these factors (and others) that allow for the colloids to be caught by the air-water interface and form a smooth deposition. While others have tested the effects of these variables before~\cite{DensityDependence,li2015coffee,dugyala2014control}, here we have shown the importance of considering all sources of coffee-ring suppression when comparing results as this suppression is controlled by many system parameters~\cite{li2015coffee, dugyala2015evaporation, PolymerCoat}. Despite these many control parameters, the width of the coffee ring is found to be very well described by equation \ref{eq:popov}, even at higher concentrations and particle aspect ratios than were considered in previous theories.

\section{Conclusions}
Here, we have demonstrated that particle anisotropy does not affect the dried deposition in a system of silica particles suspended in water. In fact, even at high concentrations, droplets formed from anisotropic particles neatly follow the predicted relationship between ring-width and concentration. The suppression of the coffee-ring effect is determined by the surface capture effect, which in turn is controlled by many parameters including particle shape, particle density, temperature~\cite{li2015coffee}, surfactant concentration~\cite{dugyala2015evaporation,carbontubessurfactants}, and the particle surface treatment~\cite{PolymerCoat}. In exploring the effect of anisotropy we observed deposit profiles consistent with those seen~\cite{li2015coffee} and predicted~\cite{2024theory} previously. We also demonstrate that characterizing the coffee-ring effect is best done using profilometry techniques rather than light microscopy, and that the width of the coffee ring is a useful and well-predicted metric for quantifying the effect control parameters have on the dried deposit.  Coffee ring formation is intrinsically linked to particle surface capture, and thus is sensitive to particle surface chemistry, substrate preparation, and the components of the suspending medium. This sensitivity makes it crucial to consider all possible sources of ring suppression when it is observed.

\section{Appendix}
\noindent\textbf{Profiles at High Humidity:} To test the effect of reducing the evaporation rate, droplets were dried at high relative humidity. RH=$75\%$ was achieved by leaving beakers of water inside the sample chamber for 4 hours prior to the experiment. We do not note substantial differences between profiles formed at RH=$10\%$ and RH$=75\%$, as depicted in Figure \ref{fig:RHprofiles}. The measured widths are plotted in Figure 3 and are consistent with those seen at all other conditions. The evaporation time at RH$=75\%$ increased to $15$ minutes, which may explain why the centers of deposits formed at high humidity appear depleted as the increased evaporation time could allow more colloids to be carried to the edge of the droplet.


\begin{acknowledgments}
This work made use of the Keck-II and the EPIC facilities of Northwestern University’s NUANCE Center, which has received support from the SHyNE Resource (NSF ECCS-2025633), the IIN, and Northwestern's MRSEC program (NSF DMR-1720139). Sam Nielsen was partially funded by the Northwestern Summer Undergraduate Research Grant. We thank Isaac Spackman for assistance in using PACKMOL to generate initial configurations for rod simulations.
\end{acknowledgments}


\bibliography{dropbib}

\end{document}